\magnification=1200\tolerance=300
\overfullrule=0pt
\hsize=11.5cm

\centerline{\bf Band ferromagnetism versus collective Kondo state}
\medskip
\centerline{{\bf in lattice fermion models}\footnote*{Presented at the 
Graduate School on Strongly Correlated Electron Systems, Debrecen (Hungary), 
September 1996. To appear in {\sl Philosophical Magazine B} (1997).}}
\vskip 1 cm
\centerline{By P. FAZEKAS}
\bigskip
\centerline{Research Institute for Solid State Physics}
\centerline{Budapest 114, P.O.B. 49, H--1525 Hungary.}

\vskip 1cm

\centerline{ABSTRACT}
\bigskip

It is becoming widely recognized that, contrary to earlier 
expectations, the 
usual one-band Hubbard model does not give an explanation for itinerant 
ferromagnetism. After reviewing the status of magnetic ordering in the 
one-band model, we discuss the possibility of ferromagnetism in some  
recently introduced two-band Hubbard models, and in generalized Anderson 
lattices. It is argued that these two classes of models are closely related 
and that it is their common feature that the ferromagnetic phase has to 
compete with a collective Kondo state.

\vskip 1cm

\centerline{\S\   1. INTRODUCTION: MAGNETISM IN THE HUBBARD MODEL}
\bigskip
Over three decades of intense research effort have been spent on mapping out  
the phase diagram of the single-orbital Hubbard model (Hubbard 1963) 
$$
{\cal H} = -t\sum_{\langle {\bf i},{\bf j}\rangle}\sum_{\sigma} 
(c_{{\bf i}\sigma}^{\dagger}c_{{\bf j}\sigma} + {\rm H.c.}) + 
Un_{{\bf i}\uparrow}n_{{\bf i}\downarrow}\; . \eqno(1)
$$
It has been usual to treat the case when the summation over {\bf i}, {\bf j} 
extends over the sites of a Bravais lattice (typically, the $d$-dimensional 
cubic lattice); then a single atomic orbital gives rise to a single tight 
binding band. In such cases, the expressions ``single-orbital'', and 
``single-band'' can be used synonymously. Recent work on carefully constructed 
non-Bravais lattices made us aware of the potential importance of 
single-orbital multi-band Hubbard models. Degenerate or multi-orbital Hubbard 
models and their extended versions 
are introduced by a completely different physical motivation, but they can be 
made formally equivalent to single-orbital models with a larger number of 
sites per unit cell.    

The extensive work on the simplest one-band case was done in the expectation 
that most of the major correlation phenomena are represented in its phase 
diagram, and are thus generic features of the Hubbard model, without the need 
for fine-tuning the parameters. Optimistically, one might have hoped to find 
Mott transition, ferro- and antiferromagnetism, incommensurate magnetic 
structures, charge and spin gaps, heavy fermions, and maybe even an 
electronic mechanism of (high-temperature) superconductivity. All seemed 
to go well when early Hartree--Fock calculations (Penn 1966) readily gave 
extensive domains of both ferro- and antiferromagnetism in the 
$n$--$U/t$ plane ($n=N/L$ is the band filling, $L$ is the number of lattice 
sites, and $N$ the number of electrons). It was taken for granted that 
allowing more complicated ordering patterns, richer Hartree--Fock phase 
diagrams can be found. But at least, there seemed to be no reason to doubt 
that a basic understanding of the most important kinds of magnetic ordering 
has been reached. One might have argued (all too naively, as it turns out) 
that for three-dimensional systems, fluctuation effects play a relatively 
minor role: they might shift the phase boundaries but would not alter the 
overall appearance of the phase diagram.  

In recent years, this view has undergone substantial changes. Now we would 
rather say that while the single-band Hubbard model provides a solid 
basis for understanding antiferromagnetic Mott insulators, and probably also 
spin density wave states, the seemingly simplest kind of magnetic order: 
ferromagnetism, is {\sl not} a generic Hubbard phenomenon. It is acknowledged 
that Nagaoka's Theorem (Nagaoka, 1966) guarantees the existence of a fully 
polarized ground state in a singular limiting case: at $U=\infty$, with one 
electron more (or less, depending on the sign of $t$) than what corresponds 
to exact half-filling. However, there seems to be no general argument which 
would permit the continuation of the Nagaoka state into an extended 
ferromagnetic region of the phase diagram (Takahashi 1982, S\"ut{\H o} 1991). 
The recently developed detailed understanding of the 
infinite-dimensional ($d=\infty$) Hubbard model also speaks against the 
possibility of ferromagnetism. A perturbation expansion of the Landau 
Fermi-liquid parameters indicates that the tendency to a ferromagnetic 
instability is strongly suppressed (M\"uller--Hartmann 1989). A recent 
version of the ground state phase diagram shows an extended antiferromagnetic 
regime straddling the $n=1$ line, flanked by SDW domains but no ferromagnetism 
(Freericks and Jarrell 1995). We have no proof which would exclude states with 
a finite polarization (apart from the Nagaoka limit) in the square lattice or 
simple cubic Hubbard models, but the region where high-spin states may still 
exist, keeps on shrinking (Wurth, Uhrig and M\"uller--Hartmann 1996), and it 
has become imaginable that these bipartite lattice models are never 
ferromagnetic at all. 
 
The (possible) elimination of one of the major Hartree--Fock phases means 
that the Hubbard model is much more fluctuating than it was originally 
assumed. The essential reason is the purely on-site nature of the interaction: 
the Hubbard term ``does not feel'' the change of the space dimensionality $d$. 
(In contrast, mean field theory would become exact at $d\to\infty$ for any 
additional intersite interaction term.) Fluctuations tend to stabilize the 
paramagnetic 
phase. It is interesting to note that a variational theory which enforces  
local correlations by Gutzwiller-projecting the optimized Hartree--Fock 
states, still finds an extended regime of (unsaturated) ferromagnetism 
(Fazekas, Menge and M\"uller--Hartmann 1990). Since the evaluation is exact 
in $d=\infty$, this finding points to the importance of correlations which 
are {\sl not} described by the Gutzwiller Ansatz.  
 
Regarding itinerant ferromagnetism as a major unsolved problem may seem 
surprising since it is well-known that the magnetic and structural properties 
of the iron-group elements, and their alloys, are described in impressive 
quantitative detail by density functional theories (Andersen, Madsen, Poulsen, 
Jepsen and Koll\'ar 1977, Gy{\H o}rffy, Koll\'ar, Pindor, Stocks, Staunton 
and Winter 1984). The difficulty arises solely in the theory of lattice 
fermion models. In lattice fermion theories, one tries to understand ordering 
phenomena by studying models with a small local basis, and a restricted set 
of coupling constants (in the Hubbard model, we have just a single orbital, 
and one coupling constant). In contrast, for density functional theories 
the fundamental quantity is the electron density, and orbitals play a 
purely auxiliary role. One needs (in principle, infinitely) many local 
basis functions to describe the spatial variation 
of the electron density.  The interaction is the true Coulomb interaction, 
with all sorts of inter-orbital matrix elements. If we wanted to translate 
a density functional calculation into a lattice fermion model reasonably 
faithfully, we would need a multi-band model with a large number of 
coupling constants. Even if such a model could be solved, and shown to be 
ferromagnetic, we might still complain about our lack of insight into the 
basic mechanism of ordering. Thus what we would really like is finding the 
``minimal'' lattice model of ferromagnetism (presumably an extension of the 
Hubbard model) with as few parameters as possible. Ferromagnetism should be 
a robust phenomenon, appearing at intermediate coupling strengths, and 
extending over a substantial range of band filling. The question is which 
is the simplest, and at the same time physically relevant, modification of 
the usual single-band Hubbard model, which would give us such results. 
Ideally, we would like a model which explains ferromagnetism at about the 
same level of ease, clarity, and also confidence, as the simple Hubbard model 
does for antiferromagnetism. The exploration of several different routes is in 
progress, and apparently, there are a number of ways to get ferromagnetism. 
We do not know yet whether they are all realized in different systems, or 
we will eventually settle for a single relevant mechanism. 

The major candidates are the extended Hubbard model (Strack and Vollhardt 
1995, Kollar, Strack and Vollhardt 1996), carefully constructed 
single-orbital models on tightly packed (usually non-Bravais) lattices 
(M\"uller--Hartmann 1995, Tasaki 1995, Penc, Shiba, Mila and Tsukagoshi 1996), 
Anderson and Kondo lattices, and double exchange models. 

\bigskip\smallskip 

{\S\  2. FERROMAGNETISM IN TWO-BAND HUBBARD MODELS}
\bigskip
One of the popular ways to get ferromagnetism in the Hubbard model is to 
consider lattices with triangular plaquettes. To understand why this idea 
works, let us begin with Tasaki's (1996) toy model of ferromagnetism: a 
single triangle. Let the sites 1 and 2 at the 
base of the triangle be connected with the hopping matrix element $-t$, while 
site 3 at the top vertex of the triangle is connected with $-v$ to 
sites 1 and 2. Furthermore, we ascribe the site energy $\epsilon_3$ to the 
top site but not to sites 1 and 2. The Hubbard $U$ may also be chosen 
differently for site 3. The three-site Hubbard model is 
$$
{\cal H}_{3} = -t\sum_{\sigma}(c_{1\sigma}^{\dagger}c_{2\sigma} + 
c_{2\sigma}^{\dagger}c_{1\sigma}) + \epsilon_3\sum_{\sigma}{\hat n}_{3\sigma}
+ U_3{\hat n}_{3\uparrow}{\hat n}_{3\downarrow} \;\;\;\;\;\;\;\;\;\;\;\;\;\;\; 
\eqno(2)
$$ 
$$
\;\;\;\;\;  -v\sum_{\sigma}
(c_{1\sigma}^{\dagger}c_{3\sigma} + c_{3\sigma}^{\dagger}c_{1\sigma} + 
c_{2\sigma}^{\dagger}c_{3\sigma} + c_{3\sigma}^{\dagger}c_{2\sigma})
+ U\sum_{j=1}^2{\hat n}_{j\uparrow}{\hat n}_{j\downarrow}\; . 
$$

Following Penc et al. (1996), we consider the strong-coupling regime 
$U, U_3\gg \epsilon_3 \gg |t|, |v|$, and take $N=2$ electrons. Then 
the low-energy configurations have one electron at each of the base sites, 
and the effective Hamiltonian can be expressed by the spins ${\bf S}_1$ and 
${\bf S}_2$. To get the exchange coupling, let us start from 
$|\uparrow\downarrow 0\rangle =c_{1\uparrow}^{\dagger}
c_{2\downarrow}^{\dagger}|0\rangle$. Then the sequence of three hopping 
events: $|\uparrow\downarrow 0\rangle\rightarrow |0\downarrow\uparrow\rangle 
\rightarrow |\downarrow 0\uparrow\rangle\rightarrow |\downarrow \uparrow 0
\rangle$ effects the exchange of the two spins. The matrix elements are 
$-v$, $-v$, and $-t$, and the two intermediate states have 
the excitation energy $\epsilon_3$. The effective Hamiltonian describes 
these third-order processes
$$
{\cal H}_{\rm eff}^{(3)} = -{{4tv^2}\over{\epsilon_3^2}}\,{\bf S}_1
\cdot{\bf S}_2 = J_3\,{\bf S}_1\cdot{\bf S}_2 \; . \eqno(3)
$$
Obviously, the sign of $t$ controls the sign of $J_3$. In particular, the 
coupling is ferromagnetic if $t>0$. 

If $U$ and $U_3$ are not kept so artificially big, there are also 
contributions from the familiar second-order processes
$$
{\cal H}_{\rm eff} = {\cal H}_{\rm eff}^{(2)} + {\cal H}_{\rm eff}^{(3)} = 
\left( {4t^2\over{U}}-{{4tv^2}\over{\epsilon_3^2}}\right)\,
{\bf S}_1\cdot{\bf S}_2 \, . \eqno(4)
$$
The overall sign of the exchange coupling is decided by the competition 
between antiferromagnetic, and ferromagnetic, pieces of the kinetic exchange.

The toy model can be extended into a quasi-one-dimensional lattice model: 
the triangle-ladder (or railroad trestle) lattice which consists of the 
parallel chains of bottom sites, and top sites. The unit cell contains 
one bottom site, and one top site. Henceforth, let us denote bottom site 
operators by $c$, and top site operators by $d$. Our previous ${\cal H}_3$ 
can be generalized to
$$ 
{\cal H} = -\sum_{j\sigma} \left[\,t(c_{j\sigma}^{\dagger}
c_{j+1,\sigma}+c_{j+1,\sigma}^{\dagger}c_{j\sigma})+
t^{\prime}(d_{j\sigma}^{\dagger}d_{j+1,\sigma}+
d_{j+1,\sigma}^{\dagger}d_{j\sigma})\right.
$$
$$
\left. +v(c_{j\sigma}^{\dagger}d_{j\sigma}+
d_{j\sigma}^{\dagger}c_{j\sigma}) +
  v^{\prime}(d_{j\sigma}^{\dagger}c_{j+1,\sigma}+
  c_{j+1,\sigma}^{\dagger}d_{j\sigma})\right]  
$$
$$
 +\epsilon \sum_{j\sigma} d_{j\sigma}^{\dagger}d_{j\sigma}
+ U_c\sum_j{\hat n}_{j\uparrow}^c{\hat n}_{j\downarrow}^c + 
U_d\sum_j{\hat n}_{j\uparrow}^d{\hat n}_{j\downarrow}^d\; , \eqno(5)
$$
where ${\hat n}_{j\uparrow}^d=d_{j\uparrow}^{\dagger}d_{j\uparrow}$, etc. 
$t^{\prime}$ is the hopping connecting nearest-neighbour top sites, and we 
distinguish $v$ acting within a lattice cell from the inter-cell $v^{\prime}$. 

Assuming again $U_c, U_d\gg \epsilon \gg {\rm all\  hopping\  amplitudes}$, 
and exact quarter filling so that the bottom chain can be filled with one 
electron per site, a systematic perturbation expansion (Penc et al. 1996) 
confirms that the effective spin-spin coupling is very similar to the toy 
model result (4)
$$
{\cal H}_{\rm eff} = \sum_j \; \left( 
{4t^2\over{U_c}}-{{4tvv^{\prime}}\over{\epsilon^2}}\right)\,
{\bf S}_j\cdot{\bf S}_{j+1} \eqno(6)
$$
where the spins are sitting on the bottom chain. We get feromagnetism if 
$U_c$ is large enough. Strictly at quarter-filling, the fully polarized 
system is an insulator because of a large single-particle gap, but it can be 
shown (Penc et al 1996) that the coupling remains ferromagnetic even if we  
move away from quarter-filling.   

The plaquette mechanism gives a very transparent reason for finding 
ferromagnetism in insulating and metallic cases alike, but at a price: 
$U$ has to be so large that the leading (second-order) kinetic exchange is 
suppressed and the usually negligible higher-order processes become 
dominating. Besides, the presence of frustration has to be assumed in the 
sense that the plaquette product $\prod t_{ij}$ must have a definite sign. 
Finally, the possibility of the perturbation expansion relies on the 
seemingly ad hoc assumption of a large $\epsilon$. But at least, it is a 
hopeful feature that the ferromagnetic order is understood to be stable 
in a finite domain of the parameter space, not only in singular limiting 
cases. Our basic hope is that ferromagnetism extends beyond the range of 
validity of the perturbational argument, and then perhaps the entire phase 
can be understood by continuation from the strong-coupling regime where we 
have a well-founded argument. There are reasons to think that indeed such is 
the case. Takahashi's (1982) numerical work on small clusters used 
$\epsilon=0$; nevertheless, it indicated that three-site exchange in 
triangular loops is instrumental in stabilizing strong ferromagnetism. Tasaki 
(1995) found that the ground state of the quarter-filled 
system is a fully saturated ferromagnet if the parameters of the Hamiltonian 
(5) satisfy a certain relationship. His argument requires that $U$ be large 
but it does not require a large $\epsilon$. Penc et al. (1996) study the 
model by a variety of techniques and find an extensive ferromagnetic phase 
which is continuously connected to the strong-coupling regime. Similar results 
were found for a related family of models by Sakamoto and Kubo (1996). Further 
evidence comes from our studies described below.
    
Starting an independent line of investigation, M\"uller--Hartmann (1995) 
studied a highly symmetrical version of (5), namely, the Hubbard chain with 
nearest- and next-nearest-neighbour hopping
$$
{\cal H} = -\sum_{j\sigma} \left[\,t_1(c_{j\sigma}^{\dagger}
c_{j+1,\sigma}+{\rm H.c.})+
t_2(c_{j\sigma}^{\dagger}c_{j+2,\sigma}+{\rm H.c.})\right] + 
U\sum_j{\hat n}_{j\uparrow}{\hat n}_{j\downarrow}\; . \eqno(7)
$$
Gauge symmetry allows to fix the sign of $t_1>0$. For $t_2>0$, the model is 
not ferromagnetic, in the sense familiar from the usual one-dimensional 
(1-dim) Hubbard model. For $t_2<0$ there is a Nagaoka state (Mattis and 
Pe${\tilde{\rm n}}$a 1974) and the question arises whether this is just an 
isolated point, or it belongs to a finite domain of ferromagnetic states. 

Noticing that the band structure develops two degenerate minima if 
$t_2/t_1<-1/4$, M\"uller--Hartmann (1995) suggested that in this case, there 
may be a ``low density route'' to ferromagnetism. Let us start with the 
observation that for $N=2$ electrons, the singlet and triplet states are 
degenerate if $U=0$, and the ground state is certainly a triplet for $U>0$. 
We may be wondering whether the high-spin state survives if we increase the 
number of electrons. M\"uller--Hartmann argued that this is the case,  
at least when $U\to\infty$, and the density is vanishingly small.

We have undertaken a detailed investigation of the extent of the 
ferromagnetically ordered state of the $t_1$--$t_2$ Hubbard chain in its  
parameter space (Pieri, Daul, Baeriswyl, Dzierzawa and Fazekas 1996; Daul, 
Pieri, Dzierzawa, Baeriswyl and Fazekas 1996). We started with exact 
diagonalization studies of small chains (with $L\le 18$ sites, $L$ and $N$ 
even, periodic boundary conditions). We found that the ground state is either 
a singlet, or fully polarized. In particular, we found that the system 
has a high-spin ground state for large enough $U$ if two conditions are 
satisfied: 1) $t_2/t_1<-1/4$, i.e., there are two degenerate band minima, 
and 2) $N$ is small enough ($N<N_{\rm cr}(t_2/t_1)$) so that the fully 
polarized Fermi sea consists of two disjoint pieces. Our data show that 
the ferromagnetic phase is quite extended, ranging from low to intermediate 
densities, and appearing at intermediate coupling strengths (Pieri at al. 
1996). This gives evidence that it is possible to get robust itinerant 
ferromagnetism in a single-orbital Hubbard model. 

There remains, however, the unsettled question of the relationship of the 
intermediate-density ferromagnetism to the Nagaoka state. Under the stated 
conditions, we did not find high-spin states for $N$ between $N_{\rm cr}$ 
and the Nagaoka value $N=L-1$. Though these (purely numerical) findings seem 
to be in nice accord with the suggestion of a hypothetical 
``valley-degeneracy-assisted'' 
ferromagnetism, we feel somewhat uneasy about their interpretation. It 
should be emphasized that, taken in themselves, the numerical studies are 
as yet inconclusive. It is thus disturbing that we have no theoretical 
argument to show that the Nagaoka state is disconnected from the 
low-to-intermediate-density ferromagnetic region; rather on the contrary. 
Relying on the experience that the ferromagnetic state, whenever it exists, 
is fully saturated, we studied its stability by the single-spin-flip 
variational Ansatz introduced by Shastry, Krishnamurty and Anderson (1990). 
The resulting phase boundary shows a sharp cusp at the critical density 
$n_{\rm cr}$ where the two Fermi lakes merge into a single Fermi sea, but the 
critical $U_{\rm cr}$ keeps on rising continuously as $n$ is increased from 
$n_{\rm cr}$ towards 1. Similar 
result is found with a more sophisticated trial state (Daul et al. 1996). 
Thus it would seem that though the critical density is likely to play  
a distinguished role in the stability criterion for ferromagnetism, it is 
not a boundary value, and the large-$U$ behaviour is ferromagnetic for all 
$0<n<1$. Further work is needed to dispel the remaining doubts.

Whatever its exact extent, the 1-dim $t_1$--$t_2$ Hubbard model 
{\sl does} have a ferromagnetic phase which can be understood by continuation 
from the low-density limit. What about higher dimensions? It would have been 
nice if postulating the existence of degenerate band minima had 
opened the way to understanding the ferromagnetism of 3-dim metals by a 
similar scenario. Alas, we found that this possibility is ruled out (Pieri 
et al. 1996), reconfirming earlier results of a similar nature (Kanamori 1963, 
Caron and Kemeny 1971). 2-dim remains a 
borderline case where ferromagnetism can arise if the density of states at 
the bottom of the band is high enough but there is no clear-cut connection  
with the presence of degenerate minima.  
\bigskip\smallskip
{\S\  3. FERROMAGNETISM IN ANDERSON LATTICES}
\bigskip
It is interesting to reinterpret the Hamiltonian (5) by postulating that 
$c$ and $d$, instead of referring to different sites in the unit cell, denote 
different orbitals at the same site. Let us assume that $U_d$ is much bigger 
than $U_c$; then (5) describes the hybridization of a strongly correlated band 
with a relatively weakly correlated band: it is an extended version of the 
well-known periodic Anderson model (PAM) which is often used to describe 
$f$-electron based heavy fermion systems (Fulde, Keller and Zwicknagl 1990). 
Compared to the standard PAM, the new features are: a finite inherent 
bandwidth for both bands; a non-zero Hubbard $U$ for both orbitals; and 
the simultaneous presence of on-site ($-v$) and inter-site ($-v^{\prime}$) 
hybridization matrix elements.   

We are thus led to posing the question about ferromagnetism in Anderson 
lattice models, and then also in the related Kondo lattice models (KLM). In 
any case, the perturbation theory argument leading to (6) is still in force, 
yielding the ferromagnetic exchange term $\sim -4tvv^{\prime}/\epsilon^2$. 
But we have plenty of indications that the 
periodic Anderson model can be ferromagnetic even with purely on-site 
hybridization (i.e., with $v^{\prime}=0$). It has long been known that 
Gutzwiller-type treatments readily give a ferromagnetic instability (Fazekas 
and Brandow 1987; Reynolds, Edwards and Hewson 1992) but taken in themselves, 
these results are probably hardly less suspect than the Hartree--Fock 
instabilities. There is, however, supportive evidence from other approaches. 
A slave boson treatment (M\"oller and W\"olfle 1993) finds a ferromagnetic 
phase above quarter-filling, in roughly the same regime where high-spin 
ground states are reported on the basis of a numerical study of the 1-dim 
PAM (Guerrero and Noack 1996). 
Turning to the KLM: a variety of techniques gave the rather astonishing result 
that the phase diagram of the 1-dim Kondo chain is largely covered by a 
ferromagnetic phase (Sigrist, Tsunetsugu, Ueda and Rice 1992; Moukouri and 
Caron 1995). We should note, however, that most of this is in the 
intermediate-to-strong-coupling regime where the KLM is not equivalent to an 
underlying PAM, so the relationship to the previously cited Anderson lattice 
results is not straightforward. The peculiarities of the 1-dim case make it 
difficult to analyse the proverbial competition between RKKY magnetism and 
Kondo states (Doniach 1977). In higher dimensions, the weak-coupling regime 
is presumably reserved for RKKY phases (including an RKKY ferromagnet at low 
band fillings), while at stronger couplings, an itinerant ferromagnet 
competes with a heavy electron liquid (Fazekas and M\"uller--Hartmann 1991). 
In any case, it is an irony of fate that the 
Hubbard model which was introduced to explain the properties of the iron 
group elements, is so reluctant to become ferromagnetic, while the Anderson 
and Kondo lattice models which were introduced to describe the often 
non-magnetic concentrated Kondo systems, have a hard time to avoid 
ferromagnetism.

We suggest that one should not hesitate to regard this observation as a clue, 
and think it over whether two-band models related to the Anderson and Kondo 
lattices are not the natural candidates to describe itinerant ferromagnetism. 
We have already indicated that there is a formal relationship between the 
PAM, and the Tasaki--M\"uller-Hartmann kind of two-band Hubbard models. 
However, the usual PAM discards $v^{\prime}$ and with that, the plaquette 
exchange term in (6), which we identified as the essential reason for 
ferromagnetism in \S\ 2. Still, the PAM seems to have a strong tendency to 
order ferromagnetically, for an as yet unclarified reason. One may suspect 
that this is reinforced by switching on $v^{\prime}$, and this leads to the 
robust order of the two-band model (5).

It is well-established that the collective singlet (Kondo) state (which, at 
exact half-filling, becomes the Kondo insulator) is one of the possible ground 
states of the PAM and the KLM. This state is characterized by a large mass 
enhancement, and a Luttinger Fermi surface (Shiba and Fazekas 1990). The 
extension of the PAM by the new terms present in (5) necessitates a 
re-investigation of the Kondo state. It is quite possible that a line of 
first-order phase transitions separates the fully developed ferromagnetic 
order from the heavy Fermi liquid. A good estimate of the ground state 
energy of the latter is a prerequisite for locating the phase boundary.
  
In a preliminary investigation (Itai and Fazekas 1996) we studied the effect 
of switching on a Hubbard $U$ for the conduction band on the Kondo energy. 
We used an extension of the well-known Gutzwiller method 
(Fazekas and Brandow 1987) to describe the non-magnetic ground state of the 
Hamiltonian (the notations now follow those customary for the PAM: the 
strongly correlated electrons are $f$-electrons, and they hybridize with a 
$d$-band)
$$
{\cal H} = \sum_{{\bf k},\sigma}\epsilon_d({\bf k})
d_{{\bf k}\sigma}^{\dagger}d_{{\bf k}\sigma}  
+ \epsilon_f\sum_{{\bf j},\sigma}{\hat n}_{{\bf j}\sigma}^f
+ U_f\sum_{\bf j}{\hat n}_{{\bf j}\uparrow}^f
{\hat n}_{{\bf j}\downarrow}^f 
$$
$$
 +U_d\sum_{\bf j}{\hat n}_{{\bf j}\uparrow}^d{\hat n}_{{\bf j}\downarrow}^d
-v\sum_{{\bf j},\sigma}(f_{{\bf j}\sigma}^{\dagger}
d_{{\bf j}\sigma} + d_{{\bf j}\sigma}^{\dagger}f_{{\bf j}\sigma}) \eqno(8)
$$
where the {\bf k} are wave vectors, and the {\bf j} are 
site indices. The $d$-bandwidth is $W$. We considered the strongly asymmetric 
Anderson model with $U_f\to\infty$ and the $f$-level 
$\epsilon_f<0$ sufficiently deep-lying so that we are in the Kondo limit: 
$1-n_f\ll 1$ where the $f$-valence is defined as 
$n_f=\langle\sum_{\sigma}{\hat n}_{{\bf j}\sigma}^f\rangle$.

As a starting point, let us recall that for the ordinary PAM, the Kondo energy 
density is given by (Fazekas and Brandow 1987)
$$
E_{\rm K}(0) = -Wn_d^0\cdot \exp{\left\{-{{\mu_0(0)-
\epsilon_f}\over{4(v^2/W)}}\right\} }\; , \eqno(9)
$$
where the subscripts 0 refer to the $v=0$ values of the corresponding 
quantities.

Our essential new result is that the extended PAM (8) has the modified Kondo 
scale (Itai and Fazekas 1996)
$$
E_{\rm K}(U_d) = -Wn_d^0\; q_d^0\cdot \exp{\left\{-{{\mu_0(U_d)-
\epsilon_f}\over{4(v^2/W)}}\right\} } \; . \eqno(10)
$$
Like (9), this holds in the weak-hybridization limit $v\ll W$.  $q_d^0$ is 
the Gutzwiller--Brinkman--Rice (Gutzwiller 1965; Brinkman and Rice 1970) band 
narrowing factor, and $\mu_0(U_d)$ is the 
interaction-depend\-ent chemical potential of the conduction band. We see 
that switching on $U_d$ changes the Kondo scale in two different ways: 
in the prefactor, and in the exponent. The prefactor describes the 
correlation-induced narrowing of the $d$-band. The exponent reflects that 
the promotion energy needed to raise an electron from the $f$-level to the 
chemical potential depends on $U_d$. 

The overall effect of switching on the conduction electron $U_d$ is the 
suppression of the lattice Kondo scale, i.e., of the binding energy of the 
singlet ground state. This can be qualitatively understood by an argument 
which is familiar from the Brinkman--Rice scenario (Brinkman and Rice 1970) 
of the Mott transition: $U_d$ suppresses polarity fluctuations in the 
$d$-orbitals, and thereby blocks the hybridization processes which, in second 
order, give rise to the effective Kondo coupling 
$J_{\rm K}\sim 4v^2/(\mu_0(U_d)-\epsilon_f)$. Though in the usual 
Gutzwillerian framework this is the expected result, it should be mentioned 
that the conclusion about the monotonically decreasing nature of 
$|E_{\rm K}(U_d)|$ is not undisputed. In a different interpretation, (8) is 
the lattice generalization of the Hamiltonian of an Anderson impurity in a 
Hubbard band; the study of the latter problem was initiated by Schork and 
Fulde (1994). While our 
result (10) is obtained from a variational trial state which can be regarded 
as the lattice version of the lowest-order Varma--Yafet state (Varma and Yafet 
1976), the relative simplicity of the impurity problem allowed pushing 
forward to higher levels of the Varma--Yafet hierarchy (Schork 1996). It was 
found that the inclusion of higher-order processes can reverse the trend, 
and give an $|E_{\rm K}(U_d)|$ which increases with $U_d$. However, relying 
on previous experience with the simple PAM, we were arguing (Fazekas and 
Itai 1996) that in the lattice case, postulating a Luttinger Fermi surface 
implies that electron--hole-type excitations have been largely taken into 
account, and the conclusion reached by using the simplest trial state is 
likely to remain essentially unaltered.
   
Having found that the Anderson--Hubbard (AH) lattice model (8) has interesting 
features, we are led to considering the similarly constructed 
Kondo--Hubbard (KH) lattice models. For a start, it is heartening to learn 
that they are ferromagnetic in a substantial range of the conduction band 
$U_d$ (Yanagisawa and Harigaya 1994). What about the Kondo scale? First, let 
us note that, in contrast to their impurity counterparts, AH and KH lattice 
models are not simply related; they represent physically different systems. In 
a KH model, the Kondo coupling between the localized ($f$) and conduction 
electron ($d$) spins is a fixed quantity, and switching on $U_d$ 
primarily leads to a more spin-polarizable electron gas, with which it is 
easier to bind into an overall singlet. We find it a plausible result that 
the lattice Kondo scale increases with $U_d$ (Shibata, Nishino, Ueda and 
Ishii 1996).     
 
Obviously, there is still a lot to do to clarify the relative stability of 
the ferromagnetic and Kondo states. There is a good chance that in the 
Anderson--Hubbard lattice model, the Kondo scale is reduced by the conduction 
electron interaction, leading to an increased domain of ferromagnetism. This 
should hold even more clearly for an extended model which includes the 
inherent $f$-bandwidth, and maybe intersite hybridization terms, so that 
ferromagnetic plaquette exchange can appear. Investigating these effects may 
well contribute to the explanation of the apparently robust  
ferromagnetism observed in some two-band Hubbard models.

\vfill\eject

\centerline{ACKNOWLEDGEMENTS}
\medskip
The author has profited greatly from discussions and/or correspondence 
with D. Baeriswyl, S. Daul, M. Dzierzawa, B. Gy{\H o}rffy, K. Itai, K. Penc, 
E. M\"uller--Hartmann, P. Pieri, H. Shiba, D. Vollhardt, and Hal Tasaki. 
Financial support by the Hungarian National Science Research Foundation 
grant OTKA T-014201 is gratefully acknowledged. 

\vskip 1cm    



\centerline{REFERENCES}

\parindent=0pt
\bigskip
ANDERSEN, O.K., MADSEN, J., POULSEN, U.K., JEPSEN, O., and KOLL\'AR, J., 1977, 
{\sl Physica}, {\bf 86}{\&}{\bf 88}B, 249.

BRINKMAN, W.F., and RICE, T.M., 1970, {\sl Phys. Rev. B}, {\bf 2}, 4302.

CARON. L.G., and KEMENY, G., 1971, {\sl Phys. Rev. B}, {\bf 4}, 150.

DAUL, S., PIERI, P., DZIERZAWA, M., BAERISWYL, D., and FAZEKAS, P., 1996, 
{\sl Physica} B, (in press).

DONIACH, S., 1977, {\sl Physica} B, {\bf 91}, 231.

FAZEKAS, P., and BRANDOW, B.H., 1987, {\sl Phys. Scripta}, {\bf 36}, 809.

FAZEKAS, P., and ITAI, K., 1996, {\sl Physica} B, (in press).

FAZEKAS, P., MENGE, B., and M\"ULLER--HARTMANN, E., 1990, {\sl Z. Phys. B -- 
Condensed Matter} {\bf 85}, 285.

FAZEKAS, P., and M\"ULLER--HARTMANN, E., 1991, {\sl Z. Phys. B -- Condensed 
Matter} {\bf 78}, 69.

FREERICKS, J.K., and JARRELL, M., 1995, {\sl Phys. Rev. Lett.} {\bf 74}, 186.

FULDE, P., KELLER, J., and ZWICKNAGL, G., 1990, {\sl Solid State Phys.}, 
{\bf 41}, 1.

GUERRERO, M. and NOACK, R.M., 1996, {\sl Phys. Rev.} B, {\bf 53}, 3707.

GUTZWILLER, M.C., 1965, {\sl Phys. Rev.}, {\bf 137}, A1726.

GY{\H O}RFFY, B.L., KOLL\'AR, J., PINDOR, A.J., STOCKS, G.M., STAUNTON, J., 
and WINTER, H., 1984, in: {\sl The Electronic Structure of Complex Systems}, 
Eds. P. PHARISEAU and W.M. TEMMERMAN, Plenum Press, New York, 593.

HUBBARD, J., 1963, {\sl Proc. Roy. Soc. (London)} A, {\bf 276}, 238.

ITAI, K., and FAZEKAS, P., 1996, {\sl Phys. Rev.} B, {\bf 54}, R752.

KANAMORI, J., 1963, {\sl Progr. Theor. Phys.}, {\bf 30}, 275.

KOLLAR, M., STRACK, R., and VOLLHARDT, D., 1996, {\sl Phys. Rev.} B, {\bf 53}, 
9225.

MATTIS, D.C., and PE${\tilde{\rm N}}$A, R.E., 1974, {\sl Phys. Rev.} B, 
{\bf 10}, 1006.

M\"OLLER, B., and W\"OLFLE, P., 1993, {\sl Phys. Rev.} B {\bf 48}, 10320.

MOUKOURI, S., and CARON, L.G., 1995, {\sl Phys. Rev.} B, {\bf 52}, R15723.

M\"ULLER--HARTMANN, E., 1989, {\sl Int. J. Modern Phys.} B, 
{\bf 5}{\&}{\bf 6}, 749; 1995, {\sl J. Low Temp. Phys.}, {\bf 99}, 349.

NAGAOKA, Y., 1966, {\sl Phys. Rev.}, {\bf 147}, 392.

PENC, K., SHIBA, H., MILA, F., and TSUKAGOSHI, T., 1996, {\sl Phys. Rev.} B, 
{\bf 54}, 4056.

PENN, D.R. 1966, {\sl Phys. Rev.}, {\bf 142}, 350.

PIERI, P., DAUL, S., BAERISWYL, D., DZIERZAWA, M., and FAZEKAS, P., 1996, 
{\sl Phys. Rev.} B, (in press).

REYNOLDS, A.M., EDWARDS, D.M., and HEWSON, A.C., 1992, {\sl J. Phys.: 
Condens. Matter}, {\bf 4}, 7589.

SAKAMOTO, H., and KUBO, K., 1996, {\sl preprint}.

SCHORK, T., 1996, {\sl Phys. Rev.} B {\bf 53}, 5626.

SCHORK, T., and FULDE, P., 1994, {\sl Phys. Rev.} B, {\bf 50}, 1345.

SHASTRY, B.S., KRISHNAMURTY, H.R., and ANDERSON, P.W., 1990, {\sl Phys. 
Rev.} B, {\bf 41}, 2375.

SHIBA, H., and FAZEKAS, P., 1990, {\sl Progr. Theor. Phys. Suppl.}, {\bf 101}, 
403. 

SHIBATA, N., NISHINO, T., UEDA, K., and ISHII, C., 1996, {\sl Phys. Rev.} B 
{\bf 53}, R8828.

SIGRIST, M., TSUNETSUGU, H., UEDA, K., and RICE, T.M., 1992, {\sl Phys. Rev.} 
B, {\bf 46}, 13838.
  
STRACK, R., and VOLLHARDT, D., 1995, {\sl J. Low Temp. Phys.}, {\bf 99}, 385.

S\"UT{\H O}, A., 1991, {\sl Commun. Math. Phys.}, {\bf 140}, 43. 

TAKAHASHI, M., 1982, {\sl J. Phys. Soc. Japan}, {\bf 51}, 3475.

TASAKI, H., 1995, {\sl Phys. Rev. Lett.}, {\bf 75}, 4678; 1996, {\sl to be 
published}.

VARMA, C.M., and YAFET, Y., 1976, {\sl Phys. Rev.} B {\bf 13}, 2950.

WURTH, P., UHRIG, G., and M\"ULLER--HARTMANN, E., 1996, {\sl Ann. Physik}, 
{\bf 5}, 148.

YANAGISAWA, T., and HARIGAYA, K., 1994, {\sl Phys. Rev.} B {\bf 50}, 9577.

\bye